\documentclass[conference]{IEEEtran}
\usepackage{cite}
\usepackage{amsmath,amssymb,amsfonts}
\usepackage{algorithmic}
\usepackage{graphicx}
\usepackage{textcomp}
\usepackage{xcolor}
\usepackage{amsmath,amssymb,amsfonts}
\usepackage{textcomp}
\usepackage{graphicx}
\usepackage{times}
\usepackage{microtype}
\usepackage[UKenglish]{babel}
\usepackage{graphicx,dblfloatfix}
\usepackage{blindtext}
\usepackage{mathtools}
\usepackage{multirow}
\usepackage{tabularx}
\usepackage{subcaption}
\usepackage{hhline}
\usepackage{arydshln}
\usepackage{listings}
\usepackage{adjustbox}
\usepackage{color,soul}
\usepackage[margin=1in]{geometry}
\begin{document}

\title{Cost-Performance Evaluation of General Compute Instances: AWS, Azure, GCP, and OCI\\

}

\author{\IEEEauthorblockN{Jay Tharwani}
\IEEEauthorblockA{Member IEEE,Independent Researcher\\
Charlotte, NC, USA\\
Email: jtharwan@alumni.uncc.edu}
\and
\IEEEauthorblockN{Arnab A Purkayastha}
\IEEEauthorblockA{Member IEEE,
Western New England University\\
Springfield, MA, USA\\
Email: arnab.purkayastha@wne.edu}

\vspace*{-1 cm}
}

\maketitle

%0.4 col

\begin{abstract}
Cloud computing has become the cornerstone of modern IT infrastructure, offering a wide range of general-purpose instances optimized for diverse workloads. This paper provides a comparative analysis of cost and performance for general-purpose compute instances across four major cloud providers: AWS, Azure, Google Cloud Platform (GCP), and Oracle Cloud Infrastructure (OCI). Using standardized configurations of 4 vCPUs and 16 GiB of RAM, the study evaluates instances based on processor architecture (Intel, AMD, ARM), pricing models, and performance benchmarks. Key findings reveal that ARM-based instances deliver superior price-performance ratios for cost-sensitive workloads, while Intel-based instances excel in enterprise-grade applications requiring versatility and reliability. The results aim to guide organizations in selecting the most cost-effective and performance-efficient cloud resources for their specific needs.

\end{abstract}
 
\vspace*{0.5 cm}

% \vspace{-1pt}
% \vspace{-5pt}
\section{Introduction}
\label{sec:Introduction}

Gone are the days where every business had its own data centre for IT infrastructure. Businesses are widely adopting the cloud for their business-critical workloads. Cloud computing has revolutionized modern IT infrastructure, enabling organizations to deploy, manage, and scale workloads with unprecedented flexibility and efficiency. Among the diverse compute options offered by cloud providers, general-purpose instances stand out for their balanced configuration of compute, memory, and networking resources. These instances are widely used for a variety of workloads, including web servers, relational databases, application hosting, and development environments.
This paper focuses on evaluating general-purpose compute instances across four leading cloud platforms: Amazon Web Services (AWS), Microsoft Azure, Google Cloud Platform (GCP), and Oracle Cloud Infrastructure (OCI). Each provider offers instance families powered by different processor architectures—Intel, AMD, and ARM—catering to diverse workload requirements and cost considerations. As the cloud landscape evolves, the need to make informed decisions about instance selection becomes critical, particularly for organizations optimizing their infrastructure for performance and cost.
The objective of this study is to conduct a comprehensive comparison of these general-purpose instances by analyzing their:

\begin{itemize}
\item Cost efficiency: Comparing hourly pricing and 1-year commitment discounts.
\item Performance metrics: Using standardized benchmarks.
\item Processor architectures: Comparing cost per-performance between manufacturers and instruction set architectures.
\end{itemize}

This analysis leverages data from official documentation, pricing calculators, and performance benchmarks. The findings aim to guide organizations in selecting the most suitable general-purpose instances, balancing cost-effectiveness with workload requirements. By providing actionable insights, this paper seeks to contribute to better decision-making in selecting cost-efficient general-purpose cloud instances.
\vspace{0.3cm}
\section{Architecture Overview: x86 (CISC) vs. ARM (RISC)}
\label{sec:archoverview}

\subsection{Overview of CISC and RISC Architectures}

\begin{itemize}
\item CISC (Complex Instruction Set Computing):
CISC architectures, such as x86, are designed to execute complex instructions in fewer lines of code. They offer a rich set of instructions and can directly perform high-level tasks. This design simplifies programming at the cost of increased hardware complexity and power consumption.

\vspace{0.5cm}

\item RISC (Reduced Instruction Set Computing): RISC architectures, like ARM, focus on executing simple instructions that take a uniform amount of time. The simplicity enables faster execution of individual instructions and lower power consumption, making RISC processors more efficient for specific workloads.\cite{dosSantosPostRISC}\cite{maag2019comparison}

 \cite{bharadwaj2022evaluation,kmiec2018comparison}
\end{itemize}

% Table placed in center for alignment

\begin{table*}[htbp]
\label{table:ciscVSrisc}
\centering
\caption{Comparison of CISC (x86) and RISC (ARM) Architectures}

\renewcommand{\arraystretch}{1.5}
\begin{tabular}{|>{\centering\arraybackslash}m{3cm}|>{\centering\arraybackslash}m{6cm}|>{\centering\arraybackslash}m{6cm}|} 
\hline
\textbf{Feature} & \textbf{CISC (x86)} & \textbf{RISC (ARM)} \\ \hline
\textbf{Instruction Set} & Rich and complex, requiring fewer lines of code. & Simplified, with fixed instruction lengths. \\ \hline
\textbf{Power Consumption} & Higher due to complexity. & Lower, making it energy-efficient. \\ \hline
\textbf{Performance per Watt} & Moderate, better for single-threaded, high-performance tasks. & Excellent, ideal for multi-threaded workloads. \\ \hline
\textbf{Programming Simplicity} & Simplifies high-level programming tasks. & Requires optimized software for peak performance. \\ \hline
\textbf{Hardware Complexity} & Higher, leading to increased power and thermal output. & Lower, enabling lightweight and efficient designs. \\ \hline
\textbf{Cost} & Typically, more expensive. & Generally cheaper due to simpler manufacturing. \\ \hline
\end{tabular}

\end{table*}

\subsection{Advantages of Each Architecture }
See Table 1 for advantages of each architecture.

\subsection{Scenarios Where Each Architecture Excels}

CISC (x86):
\begin{itemize}

\item High-Performance Applications: Suitable for compute-intensive tasks such as database management, large-scale data analytics, and virtualization, where high single-threaded performance is critical.

\item Legacy Software Support: Offers excellent backward compatibility, making it a preferred choice for applications dependent on older software.\cite{kmiec2018comparison}\cite{bharadwaj2022evaluation}

\end{itemize}

\vspace{0.5cm}
RISC (x86):
\begin{itemize}
\item Energy-Efficient Workloads: Ideal for scenarios where power efficiency and cost savings are paramount, such as web hosting, content delivery networks, and microservices.

\item Cloud-Native and Parallel Workloads: Excels in distributed systems and containerized environments due to its ability to handle multi-threaded workloads efficiently.[25][27]\cite{kmiec2018comparison}\cite{bharadwaj2022evaluation}

\end{itemize}

\subsection{Applicability in Data Centers}

CISC (x86) in Data Centers:
\begin{itemize}

\item Advantages: Dominates traditional data centers, particularly for workloads requiring consistent high performance and compatibility with legacy software.

\item Use Cases: Enterprise applications, virtual machines, high-frequency trading systems, and HPC (High-Performance Computing).\cite{aroza2012green}\cite{powA_x86_vs_arm}

\end{itemize}

\vspace{0.5cm}

RISC (ARM) in Data Centers:

\begin{itemize}
\item Advantages: Emerging as a viable alternative for energy-efficient cloud deployments. ARM processors like AWS Graviton3 provide excellent performance per watt, reducing operational costs.

\item Use Cases: Cloud-native applications, serverless computing, web hosting, and workloads with predictable patterns.[28][29]

\end{itemize}

\subsection{The Data Center Trade-Off}

The choice between x86 (CISC) and ARM (RISC) in data centers depends on specific workload requirements:

\begin{itemize}
\item If performance and compatibility are critical, x86 is the go-to architecture.

\item For cost-sensitive, energy-efficient, and scalable cloud environments, ARM is a compelling choice.\cite{blem2013power}

\end{itemize}

\section{Methodology}
\label{sec:method}
% \vspace{-10pt}

\subsection{Cloud service provider selection criteria}

The cloud service providers selected were based on cloud market share with Amazon web services leading the approximately 300-billion-dollar[19] market at 36\%. Microsoft Azure is shortly after them at 23\% and Google Cloud is shortly after at 7\%. Oracle Cloud Infrastructure was chosen as the 4th cloud provider as it is quickly gaining a lot of momentum \cite{cloud_market_share}\cite{oracle_fiscal_2025}\cite{oracle_ai_growth}\cite{sgresearch_cloud_growth}

\begin{figure}[h]
    \centering
    \includegraphics[width=\linewidth, height=0.75\linewidth]{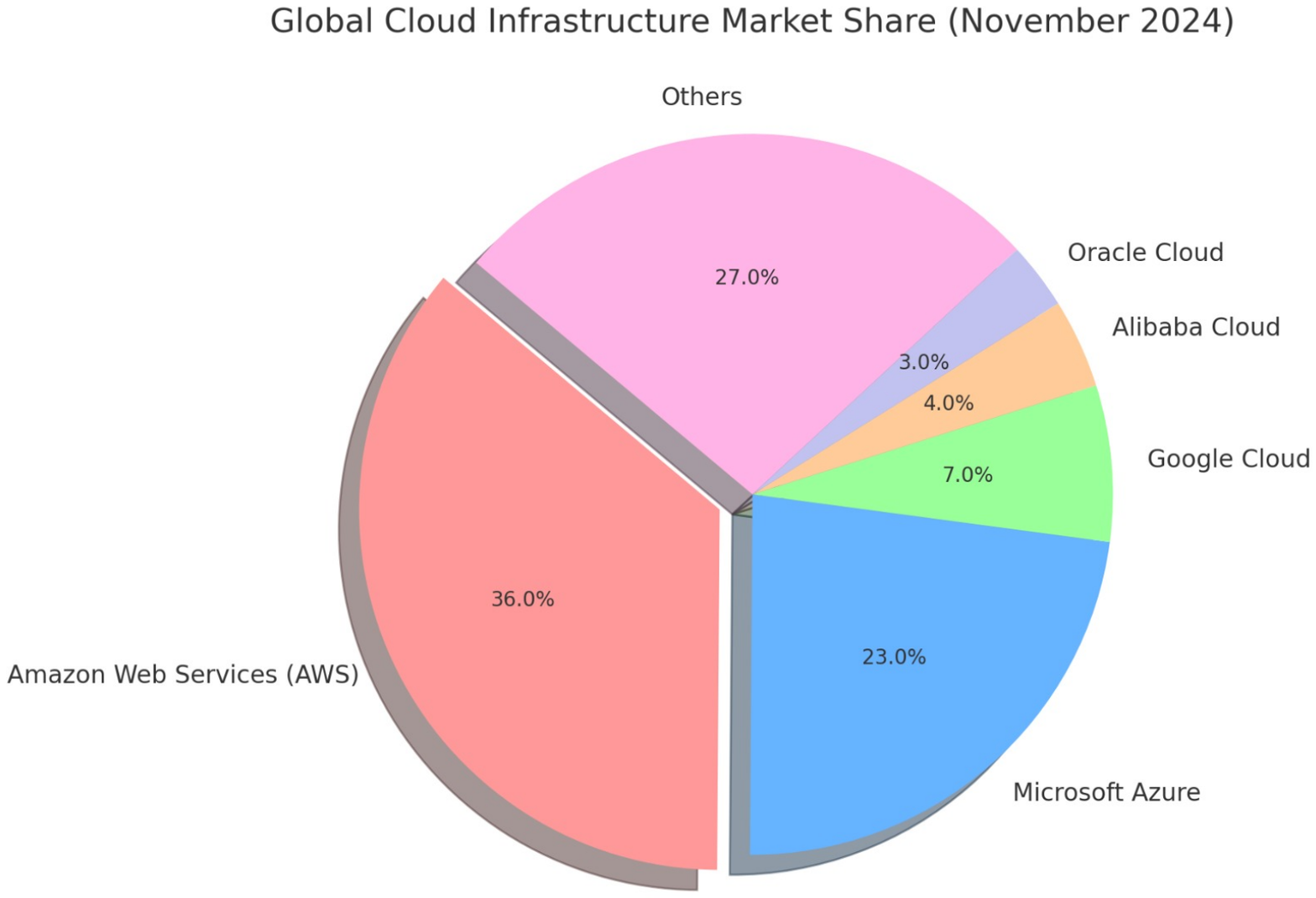}
    \vspace{-8pt}
\end{figure}

\subsection{Instance Selected}
\vspace{0.3cm}
AWS
\begin{itemize}
\item M6i: Custom Intel Xeon Platinum \cite{geekbench_m6a} 8375C (Ice Lake), x86, 4 GiB per vCPU. Use Case: Web servers, application servers, small-medium databases\cite{phoronix_aws_m6i}.

\vspace{0.5cm}

\item M6a: AMD EPYC 7R13 (Milan)\cite{azure_dv5_series}, x86, 4 GiB per vCPU. Use Case: Cost-sensitive general-purpose workloads, scalable applications.

\vspace{0.5cm}

\item M7g: AWS Graviton3 \cite{aws_graviton3}(ARM-based, Neoverse V1 cores), 4 GiB per vCPU. Use Case: Cloud-native apps, gaming servers, caching fleets, microservices.

\end{itemize}

\vspace{0.3cm}
Azure
\begin{itemize}
\item Dv5: Intel Xeon Platinum 8370C (Ice Lake)\cite{azure_dv5_series}, x86, 4 GiB per vCPU. Use Case: Enterprise applications, relational databases, web services\cite{azure_dv5_dasv5}.

\vspace{0.5cm}

\item Dasv5: AMD EPYC 7763 (Milan), x86, 4 GiB per vCPU. Use Case: Cost-efficient general-purpose tasks, database servers\cite{azure_dasv5_series}.

\vspace{0.5cm}

\item Dpsv5: Ampere Altra (80 cores @ 3.0 GHz) \cite{azure_dpsv5_dpdsv5}, ARM, 4 GiB per vCPU. Use Case: Scale-out workloads, open-source databases, modern cloud-native applications\cite{azure_dpsv5_dpdsv5}.

\end{itemize}

\vspace{0.3cm}
Google Cloud Platform (GCP)
\begin{itemize}
\item N2: Intel Xeon Platinum 8273CL (Cascade Lake), x86, 4 GiB per vCPU. Use Case: Web and application servers, enterprise apps, analytics.\cite{gcp_n2_tau_t2a}\cite{gcp_cpu_platforms}

\vspace{0.5cm}

\item N2D: AMD EPYC 7B12(Rome)[16] ,x86, 4 GiB per vCPU. Use Case: Cost-optimized general-purpose workloads, scalable applications \cite{gcp_cpu_platforms}\cite{gcp_n2_tau_t2a}.

\vspace{0.5cm}

\item Tau T2A: Ampere Altra Q64, ARM, 4 GiB per vCPU. Use Case: Cost-sensitive apps, containerized microservices, development/test environments\cite{gcp_cpu_platforms}\cite{gcp_n2_tau_t2a}.

\end{itemize}   

\vspace{0.3cm}
Oracle Cloud Infrastructure (OCI)
\begin{itemize}
\item VM Standard3.Flex : Intel Xeon Platinum 8358 (Ice Lake), x86, 4 GiB per OCPU. Use Case: Enterprise applications, dynamic web servers, small-medium databases \cite{oci_compute_shapes}.

\vspace{0.5cm}

\item VM.Standard.E4.Flex: AMD EPYC 7742 (Milan), x86, 8 GiB per OCPU. Use Case: Cost-optimized workloads, scalable general-purpose applications.\cite{oci_compute_shapes}.

\vspace{0.5cm}

\item VM.Standard.A1.Flex: Ampere Altra Q80-30, ARM, configurable memory (up to 64 GiB per core). Use Case: Cloud-native apps, portable applications, web hosting services \cite{oci_compute_shapes}.

\end{itemize}

\subsection{Instance Selection Criteria}
Among large selection of instance types between all cloud providers the instance types selected are those which are the most popular and can efficiently run general purpose workloads. These instances represent general-purpose workloads because:

\begin{enumerate}
\item Balanced Resources: They provide a consistent compute-to-memory ratio (e.g., 4 GiB per vCPU), ideal for web servers, databases, and application servers.
\item Versatility: Suitable for a wide range of workloads, from development to enterprise applications.
\item Popularity: Were actively promoted by cloud providers as default general-purpose choices in the past (e.g., AWS M6i, Azure Dv5, GCP N2, OCI VM.Standard3).
\item Diverse Architectures: Include Intel, AMD, and ARM options, catering to both traditional and modern workloads.
\item Cost-Effectiveness: Lower costs than specialized instances while maintaining performance for common use cases.
\item Global Availability: Widely available across regions, ensuring consistent performance and pricing.

\end{enumerate}

All instances amongst the 4 cloud providers were chosen to have 4 Virtual processors and 16GiB of RAM to ensure a fair comparison of performance.

\subsection{Sources of Data}

The data was collected from official cloud provider documentation, pricing calculators, and performance benchmarks. For instance, details and specifications, referred to publicly available resources on AWS, Azure, Google Cloud, and Oracle Cloud Infrastructure websites. Performance metrics were obtained from an independent benchmark, Geekbench6. Pricing information was retrieved using cloud provider calculators and websites. The instances are also standardized (no. of vCPUs and RAM) for a fair comparison\cite{aws_pricing,azure_pricing_calculator,gcp_pricing,oci_pricing}.

\subsection{Benchmark Selection}
Geekbench6 benchmarking suite was run across all instances to measure performance of the respective instances. Geekbench6\cite{geekbench6} is a cross-platform benchmark which measures a processors single-core and multi-core performance by running some tests which a lot of modern-day application might use and finally gives a Single-Core Score and a Multi-Core score depending on how well the tests performed. Tests include File Compression, Navigation, HTML5 Browser, PDF Renderer, Photo Library, Clang, Text Processing, Asset Compression, Object Detection, Background Blur, Horizon Detection, Object Remover, HDR, Photo Filter, Ray Tracer, Structure from Motion.

\subsection{Abbreviations and calculation overview}
To be space efficient the paper uses the following abbreviations in the data tables which follow this section.
\begin{itemize}
    \item CSP: This was the cloud service provider against which the instance was chosen. AWS is Amazon web services, Azure is Microsoft’s Azure, GCP is Google cloud platform and OCI is Oracle Cloud Infrastructure. 
    \item Type: This is the type of instance series of the chosen cloud provider. The instance series are typically identified by a certain type of physical processor on which the instance runs which may include its manufacturer, instruction set architecture and other parameters. Cloud providers also classify their instance series on a parameter they are optimizing, for e.g. memory optimized and there exists a lot of these series offerings. However, for this paper a general compute series has been chosen for each cloud provider. In addition, the chosen have 4 Virtual Processors and 16(GiB) of RAM.
    \item Net(Gbps): This is the max network bandwidth offered by the instance in the chosen series measured in Gigabit per second.
    \item \$/hr: Dollar per hour. This is the fees charged by the cloud service provider per hour, in USD, when using their instance. Typically, the CSPs list this data as on-demand usage on their websites 
    \item \$/hr 1 yr: Estimated dollar per hour charges when a customer commits to use the instance for a year.
    \item SCP: Single core performance score given by Geekbench, when geekbench6 was run on this instance.
    \item MCP: Multi-core performance score given by Geekbench when geekbench6 was run on this instance.
    \item MCP/\$/hr:The on-demand pricing (\$/hr) was divided by the multicore performance score to get, on demand pricing per 1 multicore performance point per hour.
    \item MCP/\$/hr 1yr:The yearlong committed pricing was divided by the multicore performance score to get a yearlong commitment pricing per 1 multicore performance point per hour
\end{itemize}

\subsection{One year commitment pricing }
This paper has used the pricing websites of various cloud providers as well as pricing calculators to estimate the per hour pricing of the cloud infrastructure when committed for a year. Typically, cloud providers offer a discount when committed to their infrastructure for a year. 

However, OCI was an outlier in this case, as they run on something called a universal credits model and the yearly pricing is not available on their website. They mention to contact the sales department and can offer an estimated discount of 20\% as compared to their on-demand pricing. This paper has kept the on-demand pricing and the one-year commitment pricing the same for OCI, due to lack of data.  However, customers may be able to get a 20\% discount when contacting their sales team.

\vspace{0.3cm}
\section{Results and Discussion}
\label{sec:results}

 \subsection{Intel-Based Instances}
 
Table \ref{table:intelInstances} shows the performance , price and price per performance scores for general purpose Intel instances\cite{aws_pricing,azure_pricing_calculator,gcp_pricing,geekbench}].

\vspace{0.3cm}
Key insights:
\begin{itemize}
\item OCI leads the price per performance score with a on demand MCP/\$/hr score of 0.0000320 as well as 1-year MCP/\$/hr score of 0.0000320.
\item If taking OCI price as a reference, the on demand MCP/\$/hr score of AWS is 81.88\% higher, Azure is 70.31\% higher and GCP is 155\% higher 
\item The 1-year commitment price is much more comparable with OCI. AWS being 12.50\% higher, Azure 16.88\% higher and GCP 60.63\% in price per performance point.
\item Overall single core and multicore performance is highest in Azure Dv5 series. However, the performance might not be a very comparable metric due to some minor changes in the underlying hardware configuration offered by each of the Cloud service providers.

\end{itemize}

\begin{table*}[htbp]
\centering
\caption{Intel based instances}
\label{table:intelInstances}

\renewcommand{\arraystretch}{1.6}
\large
\resizebox{\textwidth}{!}{
\begin{tabular}{|>{\centering\arraybackslash}m{2cm}|>{\centering\arraybackslash}m{3cm}|>{\centering\arraybackslash}m{2cm}|>{\centering\arraybackslash}m{2cm}|>{\centering\arraybackslash}m{2cm}|>{\centering\arraybackslash}m{2cm}|>{\centering\arraybackslash}m{2cm}|>{\centering\arraybackslash}m{2cm}|>{\centering\arraybackslash}m{2cm}|}
\hline
\textbf{Cloud} & \textbf{Type} & \textbf{Net (Gbps)} & \textbf{\$/hr} & \textbf{\$/hr 1 yr} & \textbf{SCP} & \textbf{MCP} & \textbf{MCP/\$/hr} & \textbf{MCP/\$/hr 1yr} \\ \hline
AWS & M6i & Up to 12 & \$0.192 & 0.1185 & 1538 & 3297 & 0.0000582 & 0.0000360 \\ \hline
Azure & Dv5 & Up to 10 & \$0.192 & 0.1317 & 1661 & 3524 & 0.0000545 & 0.0000374 \\ \hline
GCP & N2 & Up to 10 & \$0.2187 & 0.1378 & 1205 & 2682 & 0.0000816 & 0.0000514 \\ \hline
OCI & VM.Stand. 3.Flex & Up to 10 & \$0.104 & 0.104 & 1543 & 3254 & 0.0000320 & 0.0000320 \\ \hline
\end{tabular}
}

\end{table*}

\begin{table*}[htbp]
\centering
\caption{AMD Based Instances}
\label{table:amdInstances}

\renewcommand{\arraystretch}{1.6}
\large
\resizebox{\textwidth}{!}{
\begin{tabular}{|>{\centering\arraybackslash}m{2.2cm}|>{\centering\arraybackslash}m{3.5cm}|>{\centering\arraybackslash}m{2.5cm}|>{\centering\arraybackslash}m{2.2cm}|>{\centering\arraybackslash}m{2.2cm}|>{\centering\arraybackslash}m{2.2cm}|>{\centering\arraybackslash}m{2.2cm}|>{\centering\arraybackslash}m{2.5cm}|>{\centering\arraybackslash}m{2.5cm}|}
\hline
\textbf{CSP} & \textbf{Type} & \textbf{Net (Gbps)} & \textbf{\$/hr} & \textbf{\$/hr 1 yr} & \textbf{SCP} & \textbf{MCP} & \textbf{MCP/\$/hr} & \textbf{MCP/\$/hr 1yr} \\ \hline
AWS & M6a & Up to 12 & \$0.1728 & \$0.12695 & 1611 & 3641 & 0.0000475 & 0.0000349 \\ \hline
Azure & Dasv5 & Up to 10 & \$0.1720 & \$0.1175 & 1606 & 3718 & 0.0000463 & 0.0000316 \\ \hline
GCP & N2D & Up to 10 & \$0.19032 & \$0.1199 & 1561 & 3634 & 0.0000524 & 0.0000330 \\ \hline
OCI & VM.Stand.E4.Flex & Up to 10 & \$0.074 & \$0.074 & 1559 & 3588 & 0.0000206 & 0.0000206 \\ \hline
\end{tabular}
}

\end{table*}

\begin{table*}[htbp]
\centering
\caption{ARM Based Instances}
\label{table:armInstances}

\renewcommand{\arraystretch}{1.6}
\large
\resizebox{\textwidth}{!}{
\begin{tabular}{|>{\centering\arraybackslash}m{2.2cm}|>{\centering\arraybackslash}m{3.5cm}|>{\centering\arraybackslash}m{2.5cm}|>{\centering\arraybackslash}m{2.2cm}|>{\centering\arraybackslash}m{2.2cm}|>{\centering\arraybackslash}m{2.2cm}|>{\centering\arraybackslash}m{2.2cm}|>{\centering\arraybackslash}m{2.5cm}|>{\centering\arraybackslash}m{2.5cm}|}
\hline
\textbf{Cloud} & \textbf{Type} & \textbf{Net (Gbps)} & \textbf{\$/hr} & \textbf{\$/hr 1 yr} & \textbf{SCP} & \textbf{MCP} & \textbf{MCP/\$/hr} & \textbf{MCP/\$/hr 1yr} \\ \hline
AWS & M7g & Up to 12 & 0.1632 & \$0.1199(-27\%) & 1462 & 4796 & 0.0000340 & 0.0000250 \\ \hline
Azure & Dpsv5 & Up to 10 & \$0.1540 & \$0.1058 & 1108 & 3721 & 0.0000414 & 0.0000284 \\ \hline
GCP & Tau T2A & Up to 10 & \$0.154 & \$0.09702 & 1125 & 3785 & 0.0000407 & 0.0000256 \\ \hline
OCI & VM.Stand.A1.Flex & Up to 10 & \$0.064 & \$0.064 & 1122 & 3747 & 0.0000171 & 0.0000171 \\ \hline
\end{tabular}
}
\end{table*}

\subsection{AMD-Based Instances}

Table \ref{table:amdInstances} shows the performance, price and price per performance scores for general purpose AMD instances \cite{aws_pricing,azure_pricing_calculator,gcp_pricing,oci_pricing,geekbench}]. 

\vspace{0.3cm}
Key insights:

\begin{itemize}
\item OCI again leads the price for performance score with a on demand MCP/\$/hr score of 0.0000206 as well as 1-year MCP/\$/hr score of 0.0000206.
\item Taking OCI price as a reference, the on demand MCP/\$/hr score of AWS is 130.58\% higher, Azure is 124.76\% higher and GCP is 154.37\% higher 
\item The 1-year commitment price is better. AWS being 69.32\% higher, Azure 53.40\% higher and GCP 60.19\% in price per performance point.
\item Overall single core and multicore performance is comparable and highest in AWS \& Azure m6a \& Dasv5 series. However, the performance might not be a very comparable metric due to some minor changes in the underlying hardware configuration offered by each of the cloud service providers

\end{itemize}

\subsection{ARM-Based Instances}
Table 4 \ref{table:armInstances} shows the performance , price and price per performance scores for general purpose ARM instances\cite{aws_pricing,azure_pricing_calculator,gcp_pricing,oci_pricing,geekbench}].

\vspace{0.3cm}
Key insights:

\begin{itemize}
\item ARM architecture provides the least price per performance across all cloud providers. This is also is in line with the energy savings which come with the ARM architecture.
\item OCI yet again leads the price per performance score with a on demand MCP/\$/hr score of 0.0000171 as well as 1-year MCP/\$/hr score of 0.0000171.
\item If taking OCI price as a reference, the on demand MCP/\$/hr score of AWS is 98.83\% higher, Azure is 142.11\% higher and GCP is 138.01\% higher. 
\item The 1-year commitment price is much more comparable with OCI. AWS being 46.20\% higher, Azure 66.08\% higher and GCP 49.71\% higher in price per performance points.
\item Overall single core and multicore performance is significantly higher in AWS m7g series. However, the performance might not be a very comparable metric due to some minor changes in the underlying hardware configuration offered by each of the Cloud service providers.
\item AWS has its own ARM architecture CPU while the other cloud providers provide instance on Ampere Altara CPUs. 

\end{itemize}

\begin{figure*}[!htbp]
        \centering
                \includegraphics[width=.85\linewidth, height=0.65\linewidth]{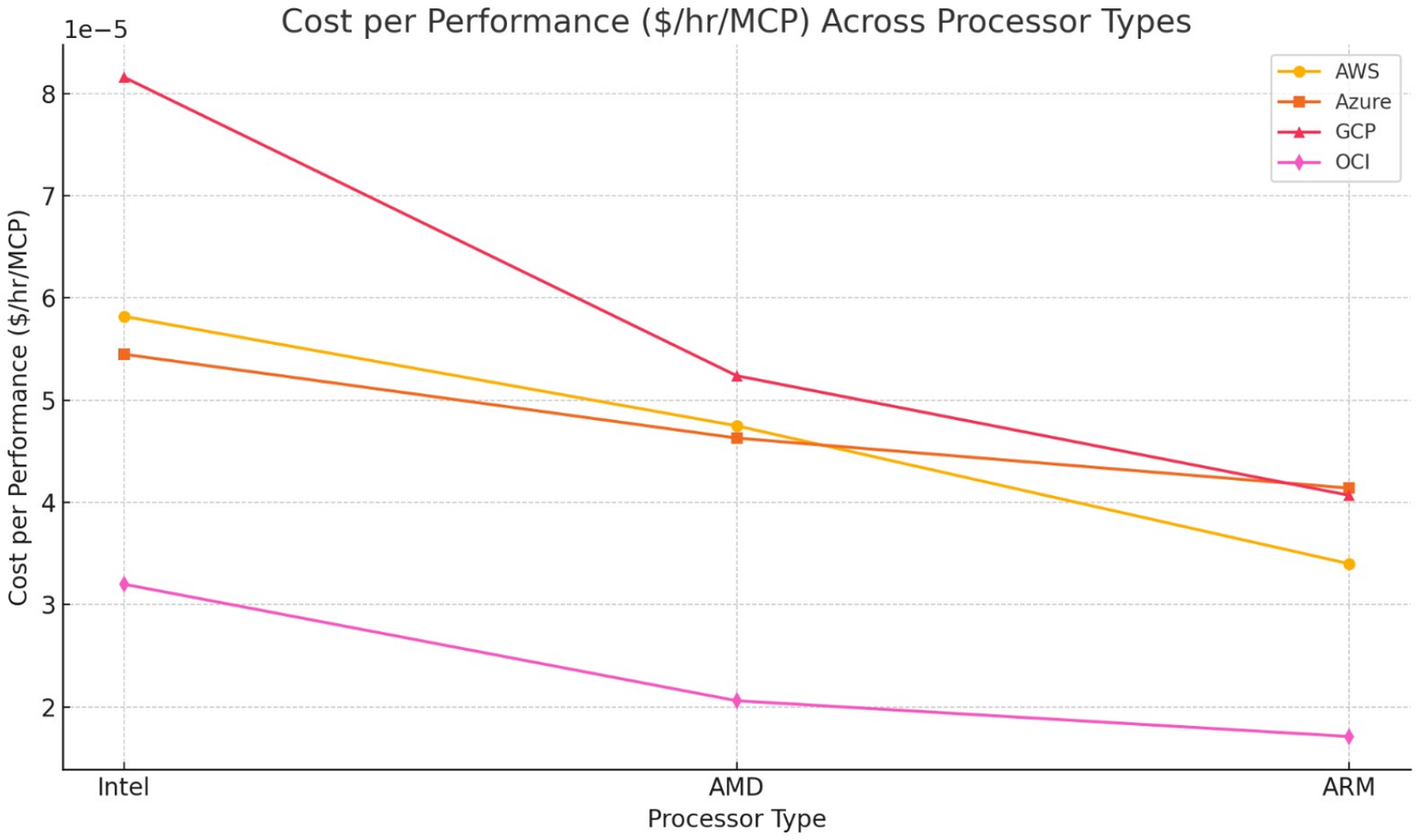}
                %\vspace{-15pt}
\end{figure*}

\begin{figure*}[]
        \centering
                \includegraphics[width=.85\linewidth, height=0.65\linewidth]{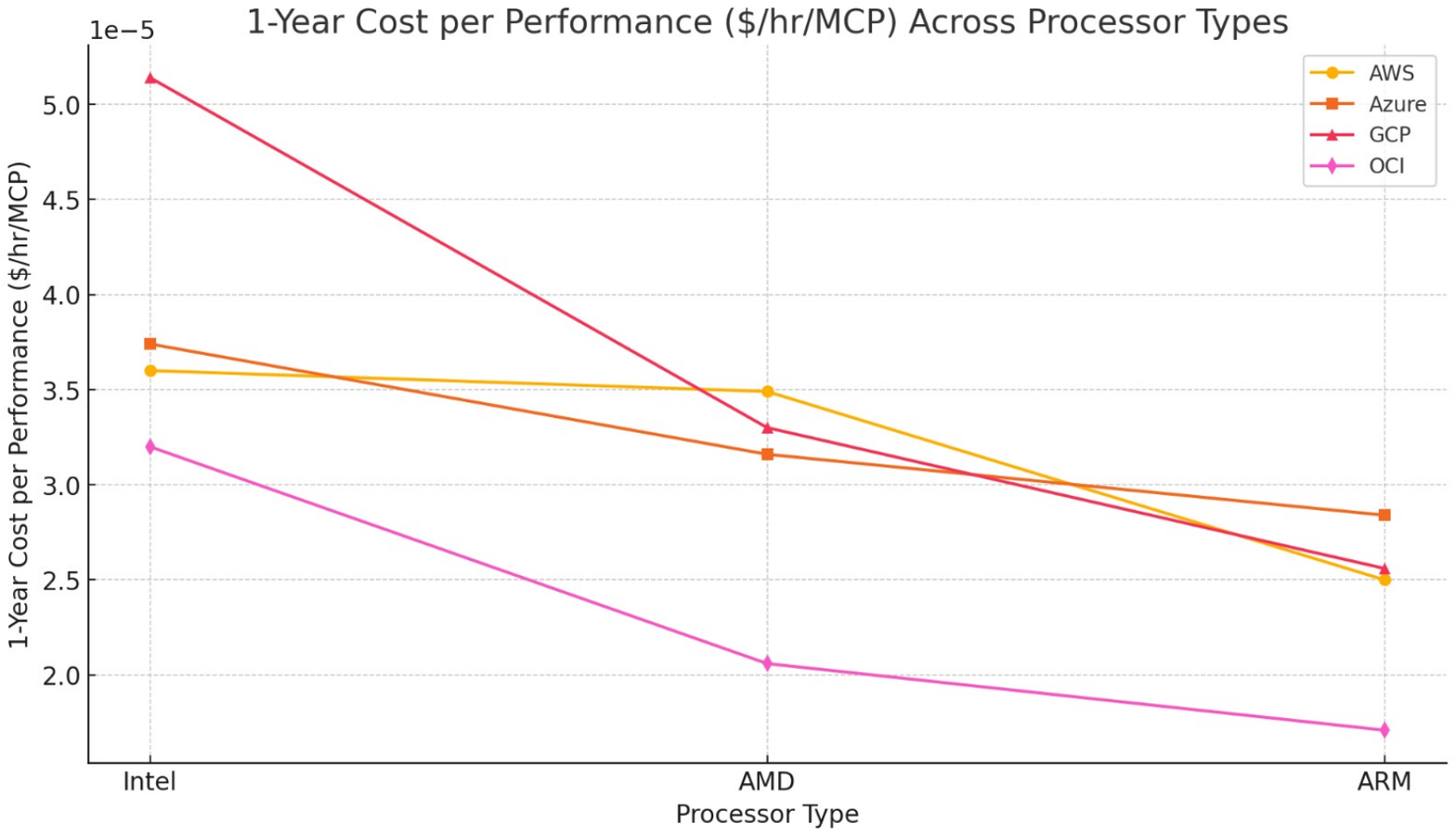}
                % \vspace{-15pt}
\end{figure*}

\begin{figure*}[]
        \centering
                \includegraphics[width=.85\linewidth, height=0.65\linewidth]{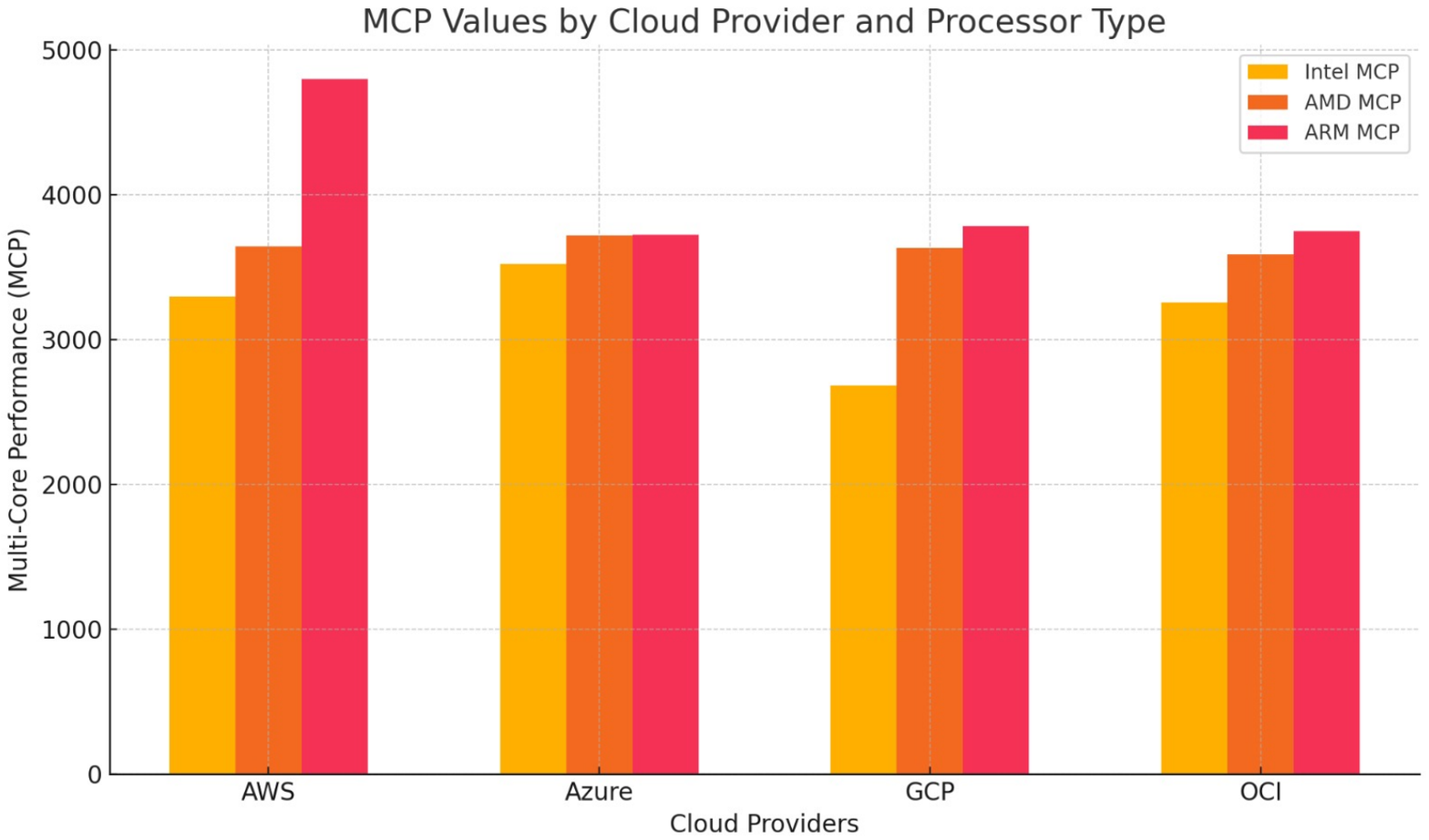}
                % \vspace{-15pt}
\end{figure*}

\subsection{Discussion}

The data reveals key trends across Intel, AMD, and  ARM-based general-purpose instances:

\begin{enumerate}
\item Performance Efficiency
    \begin{itemize}
        \item Intel (x86): AWS M6i leads with a MCP Score of 3297, but OCI VM.Standard3.Flex offers competitive performance (3254) at the lowest cost.
        \item AMD (x86): All providers deliver near-parity performance, but OCI VM.Standard.E4.Flex stands out with the lowest price-per-performance ratio.
        \item ARM: AWS M7g (4796) and GCP Tau T2A (3785) showcase ARM's growing relevance in price-performance, with OCI VM.Standard.A1 offering cost-efficient options.
    \end{itemize}
\item Cost Insights
    \begin{itemize}
        \item ARM instances consistently provide the best price-performance ratio, ideal for cost-sensitive workloads.
        \item OCI offers the lowest hourly rates across all architectures, making it appealing for budget-conscious deployments.
    \end{itemize}
\item Network Bandwidth
    \begin{itemize}
        \item Intel instances on AWS benefit from slightly higher bandwidth (12 Gbps), but parity exists across most providers and architectures.
    \end{itemize}
\item Use Cases
    \begin{itemize}
        \item Intel: Best for enterprise-grade applications and legacy software.
        \item AMD: Cost-efficient for analytics and databases.
        \item ARM: Optimal for cloud-native, containerized, and energy-efficient workloads.

    \end{itemize}
    
\end{enumerate}

\vspace{0.3cm}
\section{Future}
\label{sec:future}

This paper measures comparable general purposed compute instances across large-cap cloud providers namely AWS, Azure, GCP and an emerging cloud provider, OCI.  The paper measures money spent per CPU performance point per hour. However, the view for this paper is only limited to CPU performance. Many real-world applications might not rely only on CPU performance but also GPU performance. There are different instance types for cloud providers providing GPU support. Also, many workloads run on a cluster of virtual nodes where network bandwidth is key. 

Hence, more work can be done here where can identification of major systems of key importance can be done for different applications, assign a score to the different systems. Examples include GPU performance, memory performance and storage performance, have a mean average for these systems, assign a overall score and calculate the per hour money spent for a unit of performance score. This can help enterprises to know which cloud provider is the best choice for their money.

Even after we measure the performance for the entire system as the earlier paragraph describes, some cloud providers may still be better than others for some enterprises because of the services they provide. For eg, Oracle database services can be best provided by OCI. Hence, more work on segregating service use cases and measuring the dollar spent per performance unit can also be done.

\vspace{0.3cm}
\section{Conclusion}
\label{sec:Conclusion}

The comparative analysis of Intel, AMD, and ARM-based general-purpose instances across AWS, Azure, GCP, and OCI highlights the evolving landscape of cloud computing. ARM-based architectures, such as AWS Graviton3 and GCP Tau T2A, deliver exceptional price-performance ratios, making them a compelling choice for cost-sensitive, cloud-native workloads. AMD-based instances offer cost-efficiency for general-purpose tasks, while Intel-based instances remain the preferred option for enterprise-grade applications requiring high single-threaded performance and compatibility with legacy software.

Oracle Cloud Infrastructure (OCI) emerges as the cost leader across all architectures, particularly for AMD and ARM instances, offering an attractive proposition for budget-conscious organizations. AWS maintains a balance of performance and availability, while Azure and GCP cater to diverse workloads with competitive options.

The findings underscore the importance of aligning cloud instance selection with workload requirements. Organizations should prioritize ARM for scalability and efficiency, AMD for cost-optimized workloads, and Intel for performance-critical applications. By understanding these trade-offs, enterprises can optimize both performance and cost, ensuring their cloud infrastructure supports evolving business needs effectively.

\vspace{0.3cm}

%\vspace*{-3pt}
\bibliographystyle{IEEEtran}
\scriptsize
\bibliography{IJCTT}

\end{document}